\documentclass[preprint,onecolumn,nofootinbib]{revtex4}
\pdfoutput=1
\usepackage[colorlinks=true,linkcolor=blue,urlcolor=blue,filecolor=black,citecolor=red,pdfstartview=FitV,pdftitle={},pdfsubject={},pdfkeywords={},pdfpagemode=None,bookmarksopen=true]{hyperref}
\usepackage{graphicx}
\usepackage{amsmath}
\usepackage{amsfonts}
\usepackage{amssymb}
\usepackage{color}%
\usepackage{dcolumn}
\newcommand{\f}{\begin{equation}}
\newcommand{\ff}{\end{equation}}
\newcommand{\fa}{\begin{eqnarray}}
\newcommand{\ffa}{\end{eqnarray}}

\begin{document}
\title{Holographic response from higher derivatives with homogeneous disorder}
\author{Guoyang Fu $^{1}$}
\thanks{FuguoyangEDU@163.com}
\author{Jian-Pin Wu $^{1,2}$}
\thanks{jianpinwu@mail.bnu.edu.cn}
\author{Baicheng Xu $^{1}$}
\thanks{xubcheng@163.com}
\author{Jiaqiu Liu $^{3}$}
\affiliation{
$^1$ Institute of Gravitation and Cosmology, Department of
Physics, School of Mathematics and Physics, Bohai University, Jinzhou 121013, China\ \\
$^2$ Shanghai Key Laboratory of High Temperature Superconductors,
Shanghai, 200444, China\ \\
$^3$ College of International Exchange, Bohai University, Jinzhou 121013, China}
\begin{abstract}

In this letter, we study the charge response from higher derivatives over the background with homogeneous disorder introduced by axions.
We first explore the bounds on the higher derivatives coupling from DC conductivity and the anomalies of causality and instabilities.
Our results indicate no tighter constraints on the coupling than that over Schwarzschild-AdS (SS-AdS) background.
And then we study the optical conductivity of our holographic system.
We find that for the case with $\gamma_1<0$ and the disorder strength $\hat{\alpha}<2/\sqrt{3}$,
there is a crossover from a coherent to incoherent metallic phase as $\hat{\alpha}$ increases.
When $\hat{\alpha}$ is beyond $\hat{\alpha}=2/\sqrt{3}$ and further amplified,
a peak exhibits again at low frequency.
But it cannot be well fitted by the standard Drude formula
and new formula for describing this behavior shall be called for.
While for the holographic system with the limit of $\gamma_1\rightarrow 1/48$,
the disorder effect drives the hard-gap-like at low frequency into
the soft gap and suppresses the pronounced peak at medium frequency.

\end{abstract}
\maketitle
\section{Introduction}

The quantum critical (QC) dynamics described by CFT
or its proximity physics is strongly coupled systems without quasi-particles descriptions \cite{Sachdev:QPT} (also refer to \cite{Cha:1991,Damle:1997rxu,Smakov:2005,WitczakKrempa:2012um,Chen:2013ppa,Gazit:2013hga,Gazit:2014,Witczak-Krempa:2015jca,Lucas:2016fju}).
In dealing with these problems,
a controlled manner in traditional field theory can not usually be performed.
The AdS/CFT correspondence \cite{Maldacena:1997re,Gubser:1998bc,Witten:1998qj,Aharony:1999ti}
provides valuable lessons to understand such systems by
mapping certain CFTs to higher dimensional classical gravity.
Studying the optical conductivity $\sigma(\omega/T)$ by introducing
the probe Maxwell field coupled to the Weyl tensor $C_{\mu\nu\rho\sigma}$
in the Schwarzschild-AdS (SS-AdS) black brane background,
we find that the behavior of conductivity is similar with one in the superfluid-insulator
quantum critical point (QCP) described by the boson Hubbard model \cite{Myers:2010pk,Sachdev:2011wg,Hartnoll:2016apf}.
It provides possible route to access this kind of problems.
Further, to test the robustness of higher derivative (HD) terms,
the author of Ref.\cite{Witczak-Krempa:2013aea} studies the charge response
of a large class of allowed HD terms in the SS-AdS geometry
and find some interesting results.
Of particular interest is that the optical conductivity displays an arbitrarily sharp Drude-like peak
and the bounds found in \cite{Myers:2010pk,Ritz:2008kh} are violated.

Also, we explore the charge transport with Weyl term in a specific thermal state
with homogeneous disorder,
which is introduced by a pair of spatial linear dependent axionic fields in AdS geometry
and is away from quantum critical point (QCP), and new physics is qualitatively found \cite{Wu:2016jjd}.
Of particular interesting is that for the positive Weyl coupling parameter $\gamma>0$,
the strong homogeneous disorder drives the Drude-like peak in QCP state described by Maxwell-Weyl system in SS-AdS geometry \cite{Myers:2010pk}
into the incoherent metallic state with a dip, which is away from QCP.
While an oppositive scenario is found for $\gamma<0$.
In addition, the particle-vortex duality in the dual field theory
induced by the bulk electromagnetic (EM) duality related by changing the sign of $\gamma$
is still preserved. Nonetheless, there is still bound for the conductivity as in \cite{Myers:2010pk,Ritz:2008kh}.
In this letter, we study the charge response of a large class of HD terms
in the specific thermal state with homogeneous disorder.
The letter is organized as follows.
We describe the holographic setup for a class of HD theory with homogeneous disorder in Section \ref{sec-setup}.
And then the constraints imposing on the HD coupling parameters in the Einstein-axions-AdS (EA-AdS) geometry are explored in Section \ref{sec-bound}.
In Section \ref{sec-optical-conductivity}, we mainly study the optical conductivity from HD theory with homogeneous disorder.
Finally the conclusion and discussion are presented in Section \ref{sec-conclusion}.

\section{Holographic setup}\label{sec-setup}

A specific thermal excited state with homogeneous disorder
can be holographically described by the EA theory \cite{Andrade:2013gsa},
\fa
\label{ac-ax}
S_0=\int d^4x\sqrt{-g}\Big(R+6-\frac{1}{2}\sum_{I=x,y}(\partial \phi_I)^2\Big)
\,,
\ffa
where $\phi_I=\alpha x_I$ with $I=x,y$ and $\alpha$ being a constant.
In the action above, there is a negative cosmological constant $\Lambda=-6$,
which supports an asymptotically AdS spacetimes \footnote{We have set the AdS radius $L=1$ without loss of generality.}.
The EA action (\ref{ac-ax}) gives
a neutral black brane solution \cite{Andrade:2013gsa}
\fa
\label{bl-br}
ds^2=\frac{1}{u^2}\Big(-f(u)dt^2+\frac{1}{f(u)}du^2+dx^2+dy^2\Big)\,,
\ffa
where
\fa
\label{fu}
f(u)=(1-u)p(u)\,,~~~~~~~
p(u)=\frac{\sqrt{1+6\hat{\alpha}^2}-2\hat{\alpha}^2-1}{\hat{\alpha}^2}u^2+u+1\,.
\ffa
$u=0$ is the asymptotically AdS boundary while the horizon locates at $u=1$.
Note that we have parameterized the black brane solution by $\hat{\alpha}=\alpha/4\pi T$
with the Hawking temperature $T=p(1)/4\pi$.
Although the momentum dissipates due to the break of microscopic translational symmetry,
the geometry is homogeneous and so we refer to this mechanism as homogeneous disorder
and $\hat{\alpha}$ denotes the strength of disorder \footnote{The other models,
for instance \cite{Grozdanov:2015qia,Donos:2013eha,Donos:2014uba,Ling:2015exa,Ling:2015epa,Vegh:2013sk,Blake:2013bqa,Donos:2012js,Kim:2014bza,Davison:2014lua,Gouteraux:2016wxj,Ling:2016lis},
have also been developed to produce the effect of homogeneous disorder.}.

Now, we consider the following action beyond Weyl
\cite{Witczak-Krempa:2013aea}
\fa
\label{ac-SA}
S_A=\int d^4x\sqrt{-g}\Big(-\frac{1}{8g_F^2}F_{\mu\nu}X^{\mu\nu\rho\sigma}F_{\rho\sigma}\Big)\,,
\ffa
where the tensor $X$ is an infinite family of HD terms
\fa
X_{\mu\nu}^{\ \ \rho\sigma}&=&
I_{\mu\nu}^{\ \ \rho\sigma}-8\gamma_{1,1} C_{\mu\nu}^{\ \ \rho\sigma}
-4\gamma_{2,1}C^2I_{\mu\nu}^{\ \ \rho\sigma}
-8\gamma_{2,2}C_{\mu\nu}^{\ \ \alpha\beta}C_{\alpha\beta}^{\ \ \rho\sigma}
\nonumber
\\
&&
-4\gamma_{3,1}C^3I_{\mu\nu}^{\ \ \rho\sigma}
-8\gamma_{3,2}C^2C_{\mu\nu}^{\ \ \rho\sigma}
-8\gamma_{3,3}C_{\mu\nu}^{\ \ \alpha_1\beta_1}C_{\alpha_1\beta_1}^{\ \ \ \alpha_2\beta_2}C_{\alpha_2\beta_2}^{\ \ \ \rho\sigma}
+\ldots
\,.
\label{X-tensor}
\ffa
In the equation above,
$I_{\mu\nu}^{\ \ \rho\sigma}=\delta_{\mu}^{\ \rho}\delta_{\nu}^{\ \sigma}-\delta_{\mu}^{\ \sigma}\delta_{\nu}^{\ \rho}$
is an identity matrix and $C^n=C_{\mu\nu}^{\ \ \alpha_1\beta_1}C_{\alpha_1\beta_1}^{\ \ \ \alpha_2\beta_2}\ldots C_{\alpha_{n-1}\beta_{n-1}}^{\ \ \ \ \ \ \ \mu\nu}$.
$g_F^2$ is an effective dimensionless gauge coupling and
we set $g_F=1$ in the numerical calculation.
The action (\ref{ac-SA}) is constructed in terms of double EM field strengths,
which is sufficient for linear response,
coupled to any number of symmetry-allowed curvature tensors,
which go beyond the Weyl action studied in \cite{Myers:2010pk,WitczakKrempa:2012gn,WitczakKrempa:2013ht,Witczak-Krempa:2013nua,Katz:2014rla,Bai:2013tfa,Ritz:2008kh,Wu:2016jjd,
Wu:2010vr,Ma:2011zze,Momeni:2011ca,Momeni:2012ab,Zhao:2012kp,Momeni:2013fma,Momeni:2014efa,Zhang:2015eea,Mansoori:2016zbp,Ling:2016dck}.
Note that the $X$ tensor possess the following symmetries
\fa
X_{\mu\nu\rho\sigma}=X_{[\mu\nu][\rho\sigma]}=X_{\rho\sigma\mu\nu}\,.
\label{X-sym}
\ffa
When we set $X_{\mu\nu}^{\ \ \rho\sigma}=I_{\mu\nu}^{\ \ \rho\sigma}$, the theory (\ref{ac-SA}) reduces to the standard Maxwell theory.

Since the action (\ref{ac-SA}) is an infinite family of $n$ powers of the Weyl tensor $C$,
we truncate it to the level $n=2$, which is the $6$ derivatives and the focus in this paper.
On the other hand, since the effect of the coupling terms $\gamma_1$ and $\gamma_2$ is similar,
we mainly focus on the term of $\gamma_1$ through this paper.
For convenience, we denote $\gamma_{1,1}=\gamma$ and $\gamma_{2,i}=\gamma_i (i=1,2)$ in what follows.

\section{Bounds on the coupling}\label{sec-bound}

In this section, we examine the bounds on the coupling over the EA-AdS background (\ref{bl-br}).
To this end, we turn on the perturbations of gauge field and write down the corresponding linearized perturbative equations in momentum space \cite{Myers:2010pk,Wu:2016jjd}
\fa
&&
A'''_t+\Big(\frac{f'}{f}-\frac{X'_1}{X_1}+2\frac{X'_3}{X_3}\Big)A''_t
+\Big(-\frac{\mathfrak{p}^2\hat{q}^2X_1}{fX_3}+\frac{\mathfrak{p}^2\hat{\omega}^2X_1}{f^2X_5}+\frac{f'X'_3}{fX_3}-\frac{X_1'X_3'}{X_1X_3}+\frac{X_3''}{X_3}\Big)A'_t=0\,,
\nonumber
\\
&&
A''_y
+\Big(\frac{f'}{f}+\frac{X'_6}{X_6}\Big)A'_y
+\frac{\mathfrak{p}^2}{f^2}\Big(\hat{\omega}^2\frac{X_2}{X_6}-\hat{q}^2f\frac{X_4}{X_6}\Big)A_y
=0\,,
\label{Ma-Ay}
\ffa
where $\bf{\hat{q}}^{\mu}$$=(\hat{\omega},\hat{q},0)$ are the dimensionless quantities defined as
\fa
\hat{\omega}\equiv\frac{\omega}{4\pi T}=\frac{\omega}{\mathfrak{p}}\,,
~~~~~
\hat{q}\equiv\frac{q}{4\pi T}=\frac{q}{\mathfrak{p}}\,,
~~~~~
\mathfrak{p}\equiv p(1)=4\pi T\,.
\label{hat-omega-q}
\ffa
$X_i,\, i=1,\ldots,6$, are the matrix elements of $X_{A}^{\ B}=\text{diag}(X_i(u))$ with $A,B\in\{tx,ty,tu,xy,xu,yu\}$,
which encode the essential information of tensor $X_{\mu\nu}^{\ \ \rho\sigma}$.
In particular, $X_1(u)=X_2(u)=X_5(u)=X_6(u)$ and $X_3(u)=X_4(u)$ because of the symmetry of the background geometry (\ref{bl-br})
and the structure of $X$ tensor (\ref{X-tensor}).
Note that here we only consider the equations for $A_t$ and $A_y$ since $A'_t=-\frac{\hat{q}f}{\hat{\omega}}\frac{X_5}{X_3}A'_x$
and we have chosen the radial gauge as $A_{u}=0$.
In addition, the equations of EM dual theory can be obtained by letting $X_i\rightarrow\widehat{X}_i=1/X_i$.

\subsection{Bounds from DC conductivity}

In \cite{Witczak-Krempa:2013aea}, it has been found that for the SS-AdS geometry and the subspace of $\gamma_1\neq 0$
but other parameters vanishing,
$\gamma_1$ is unconstrained from the anomalies of causality and instabilities.
But an additional constraint from $Re\sigma(\omega)\geq 0$, especially DC conductivity being positive,
gives $\gamma_1\leq 1/48$.
Here we shall further examine this bound over EA-AdS geometry from conductivity.

\begin{figure}
\center{
\includegraphics[scale=0.7]{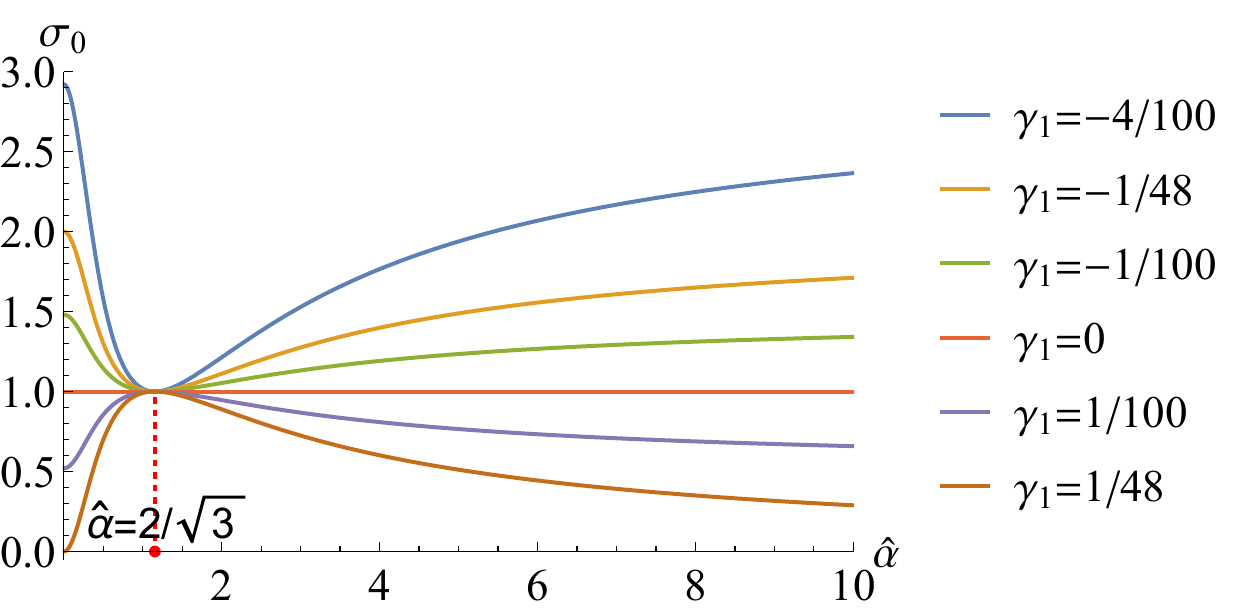}\ \\
\caption{\label{fig-dc} The DC conductivity $\sigma_0$ versus $\hat{\alpha}$ for some fixed $\gamma_1$ and other coupling parameters vanishing.
}}
\end{figure}

To this end, we write down the expression of DC conductivity \cite{Myers:2010pk,Ritz:2008kh,Wu:2016jjd}
\fa
\sigma_0=\sqrt{-g}g^{xx}\sqrt{-g^{tt}g^{uu}X_1X_5}\mid_{u=1}\,.
\label{DC}
\ffa
It can be explicitly worked out as up to $6$ derivatives
\fa
\sigma_0=1-\frac{2}{3}\gamma f''(1)-\Big(\frac{4}{3}\gamma_1+\frac{1}{9}\gamma_2\Big)f''(1)^2\,,
\label{DCv1}
\ffa
with
\fa
f''(1)=-2-\frac{4(-1-2\hat{\alpha}^2+\sqrt{1+6\hat{\alpha}^2})}{\hat{\alpha}^2}\,.
\ffa
We can easily find that for arbitrary $\hat{\alpha}$,
the upper bound $\gamma_1=1/48$ preserves for other parameters vanishing.
FIG.\ref{fig-dc} clearly shows this result.

Before closing this subsection, we would like to present some comments on the DC conductivity $\sigma_0$ from $6$ derivative term.
First, similar with the case of $4$ derivative term \cite{Wu:2016jjd}, there is a specific value $\hat{\alpha}=2/\sqrt{3}$,
for which $\sigma_0$ is independent of the coupling parameter $\gamma_1$.
Second, for the case of $4$ derivatives, $\sigma_0$ is monotonic function of $\hat{\alpha}$ \cite{Wu:2016jjd},
while for one of $6$ derivatives, $\sigma_0$ is non-monotonic and has upper/lower bound setting by $\sigma_0=1$.

\subsection{Bounds from anomalies of causality and instabilities}

To examine the causality and instabilities,
we package the equations (\ref{Ma-Ay}) into a more convenient Schr\"odinger form,
\fa
-\partial_z^2\psi_i(z)+V_i(u)\psi_i(z)=\hat{\omega}^2\psi_i(z)\,,
\label{Sch-form}
\ffa
where $dz/du=\mathfrak{p}/f$.
Note that in order to remove linear terms of $\partial_z$ and obtain the above equation,
we have introduced an auxiliary functions $G_i(u)$ to write $A_i(u)=G_i(u)\psi_i(u)$ with
$A_{\bar{t}}(u):=A'_t(u)$ and $i=\bar{t},y$.
In addition, $V_i(u)$ are the effective potentials, which are decomposed into the momentum dependent part and the independent one
\fa
V_i(u)=\hat{q}^2V_{0i}(u)+V_{1i}(u)\,,
\label{Vi-V0-V1}
\ffa
where
\fa
&&
V_{0\bar{t}}=f\frac{X_1}{X_3}\,,
\,\,\,\,\,\,\,\,\,\,
V_{0y}=f\frac{X_3}{X_1}\,,
\
\\
&&
V_{1\bar{t}}=\frac{f}{4\mathfrak{p}^2X_1^2}[3f(X_1')^2-2X_1(fX_1')']\,,
\
\\
&&
V_{1y}=\frac{f}{4\mathfrak{p}^2X_1^2}[-f(X_1')^2+2X_1(fX_1')']\,.
\ffa
Since $V_{\bar{t}}=V_y|_{X_i\rightarrow \widehat{X}_i}$,
we only consider the case of $V_{\bar{t}}$ in what follows.

\begin{figure}
\center{
\includegraphics[scale=0.6]{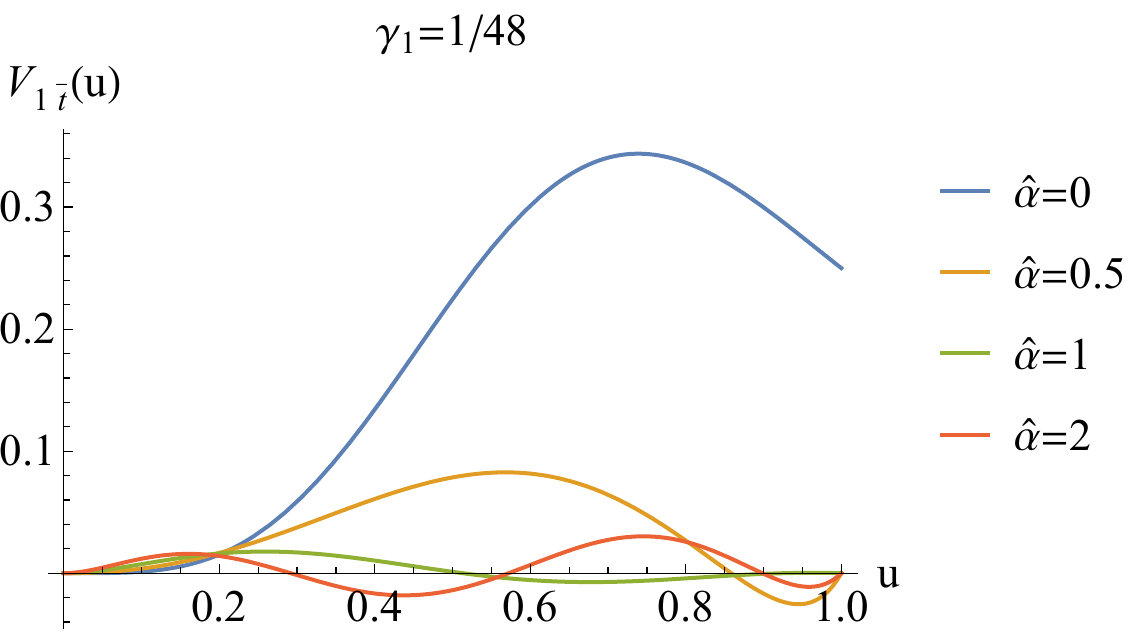}\ \hspace{0.8cm}
\includegraphics[scale=0.6]{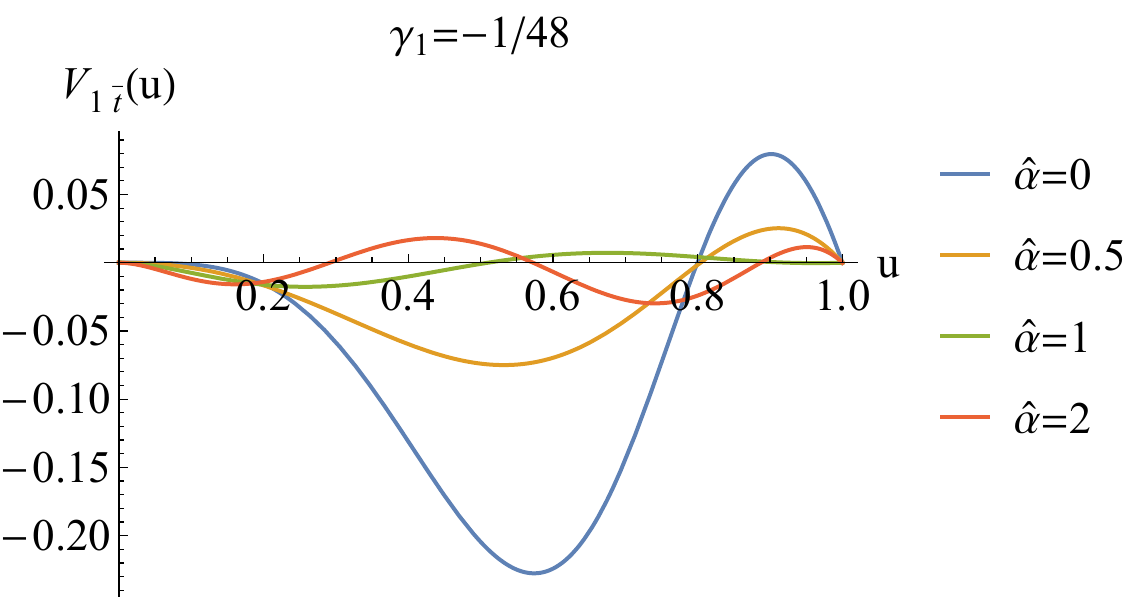}\ \\
\includegraphics[scale=0.55]{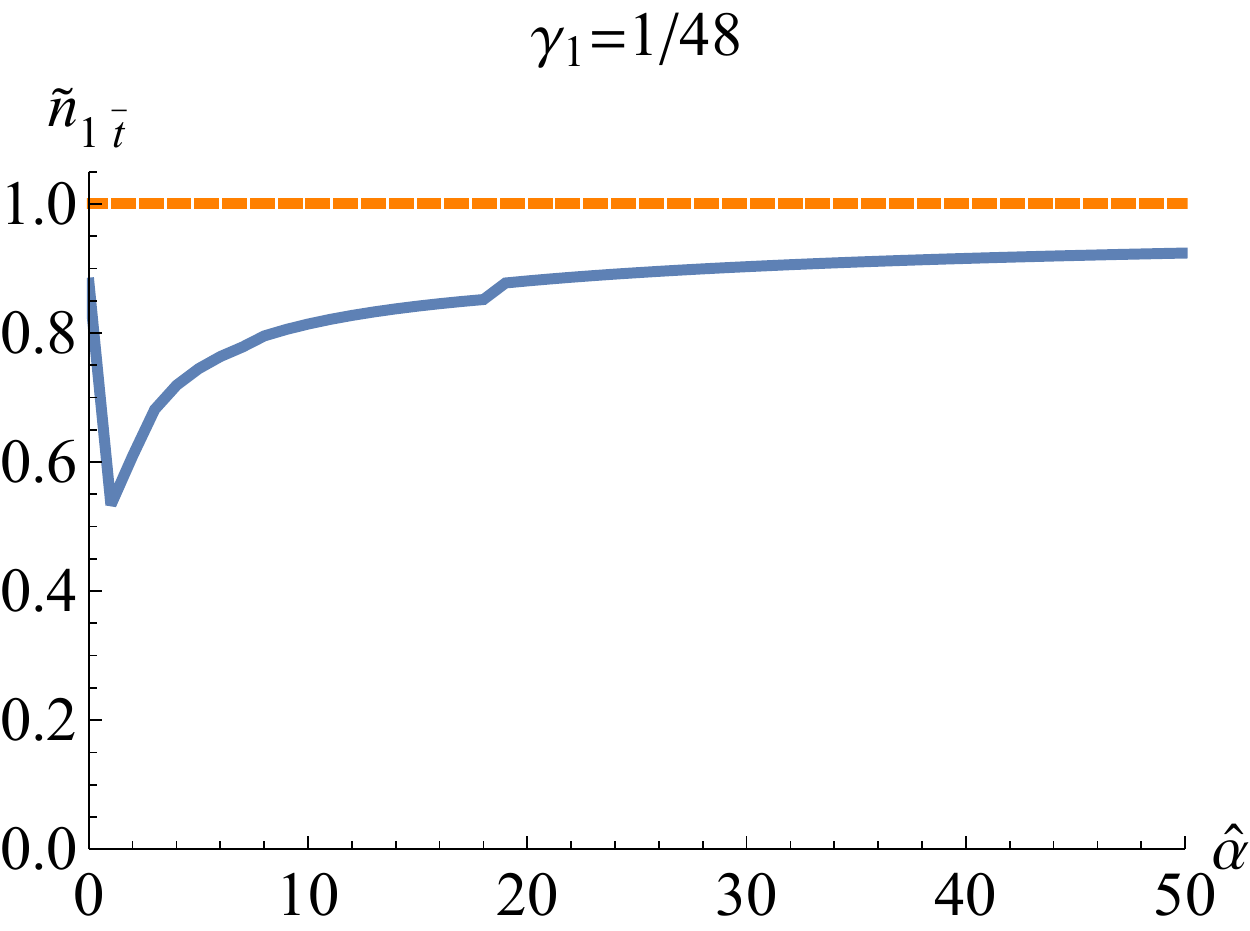}\ \hspace{0.8cm}
\includegraphics[scale=0.55]{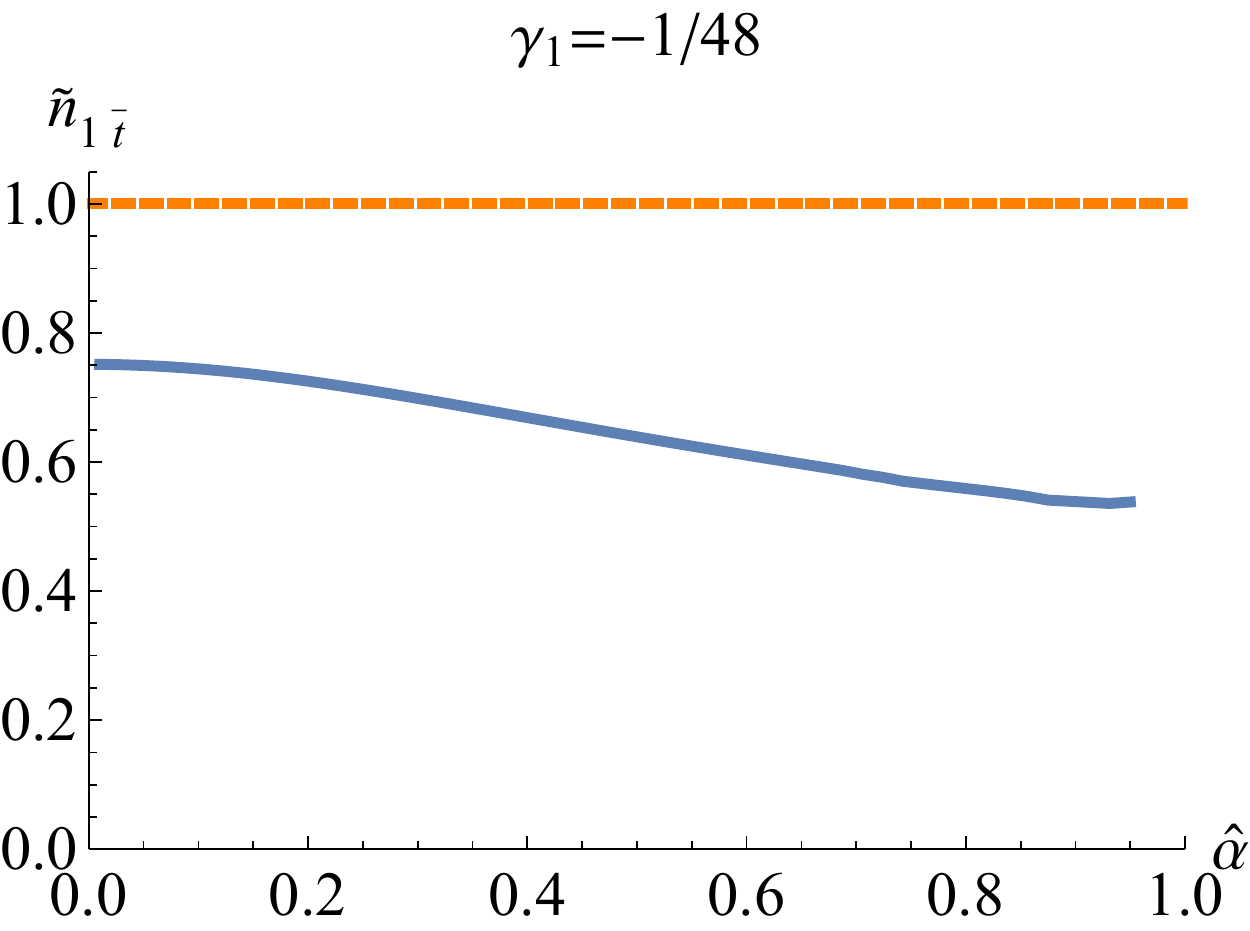}\ \\
\caption{\label{fig-v1} Top plots: The potential $V_{1\bar{t}}$ for sample $\gamma_1$ and $\hat{\alpha}$.
Bottom plots: $\tilde{n}_{1\bar{t}}$ as the function $\hat{\alpha}$ for given $\gamma_1$.
They clearly exhibit that $\tilde{n}_{1\bar{t}}$ is always less than unit.}}
\end{figure}

Now we consider the subspace of $\gamma_1\neq 0$ but other parameters vanishing.
Since $X_1=X_3$ in this case,
one has $V_{0\bar{t}}=f(u)$, which satisfies the constraint in the limit of large momentum ($\hat{q}\rightarrow\infty$),
\fa
0\leq V_{0i}(u)\leq 1\,,
\label{V0i-constraint}
\ffa
for any $\gamma_1$ and $\hat{\alpha}$.
While for the small momentum region, although the dominative potential $V_{1\bar{t}}$
develops a negative minimum (see top plots in FIG.\ref{fig-v1}),
there are no unstable modes in these regions by analyzing the zero energy bound state in the potential $V_{1\bar{t}}$ (see bottom plots in FIG.\ref{fig-v1}),
which is \cite{Myers:2007we}
\fa
\label{zero-erergy-bound}
\tilde{n}_{1\bar{t}}=I/\pi+1/2\,,\,\,\,\,\,\,\,
I\equiv
\Big(n-\frac{1}{2}\Big)\pi
=\int_{u_0}^{u_1}\frac{\mathfrak{p}}{f(u)}\sqrt{-V_{1\bar{t}}(u)}du
\,,
\ffa
where $n$ is a positive integer and the interval $[u_0,u_1]$ defines the negative well.

\begin{figure}
\center{
\includegraphics[scale=0.6]{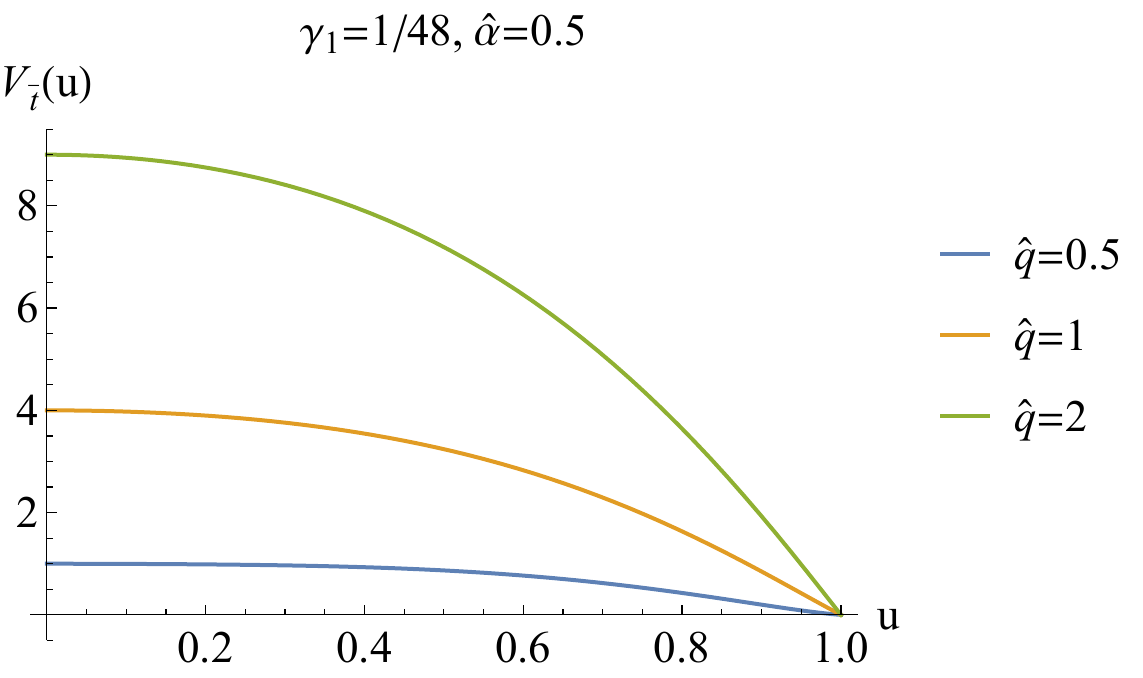}\ \hspace{0.8cm}
\includegraphics[scale=0.6]{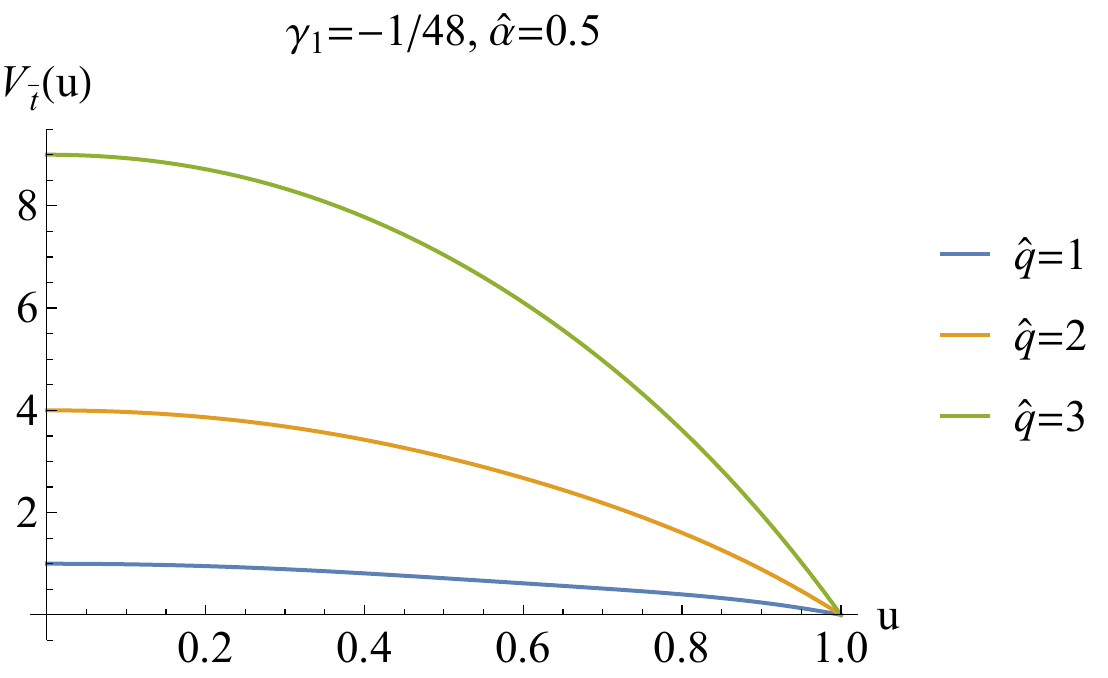}\ \\
\includegraphics[scale=0.61]{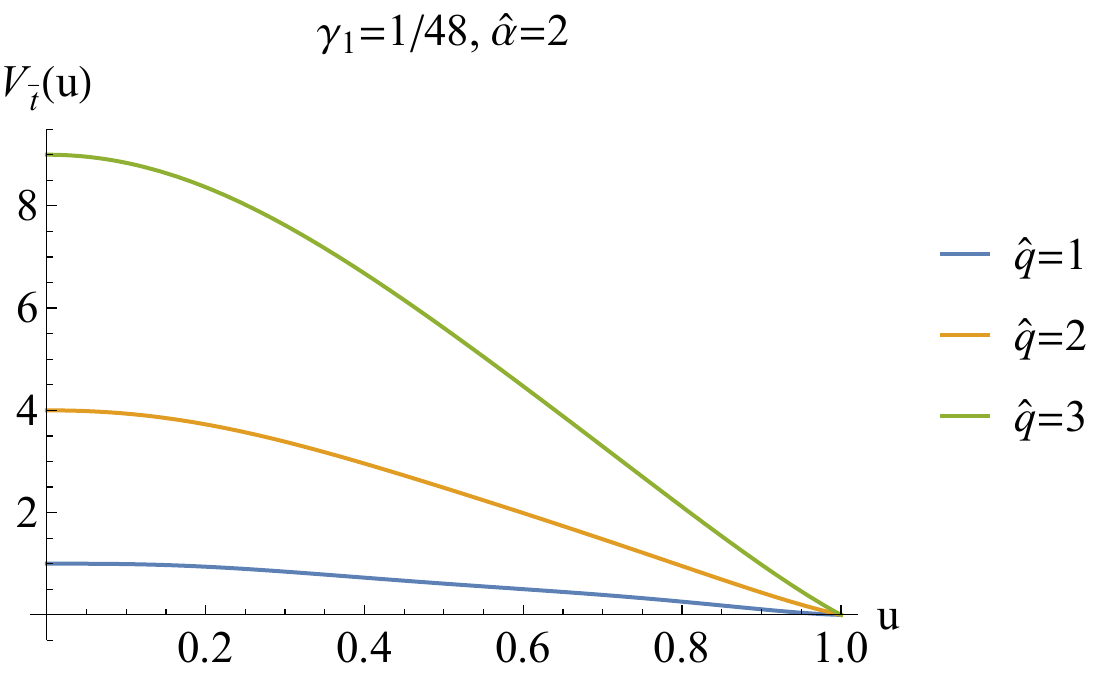}\ \hspace{0.8cm}
\includegraphics[scale=0.61]{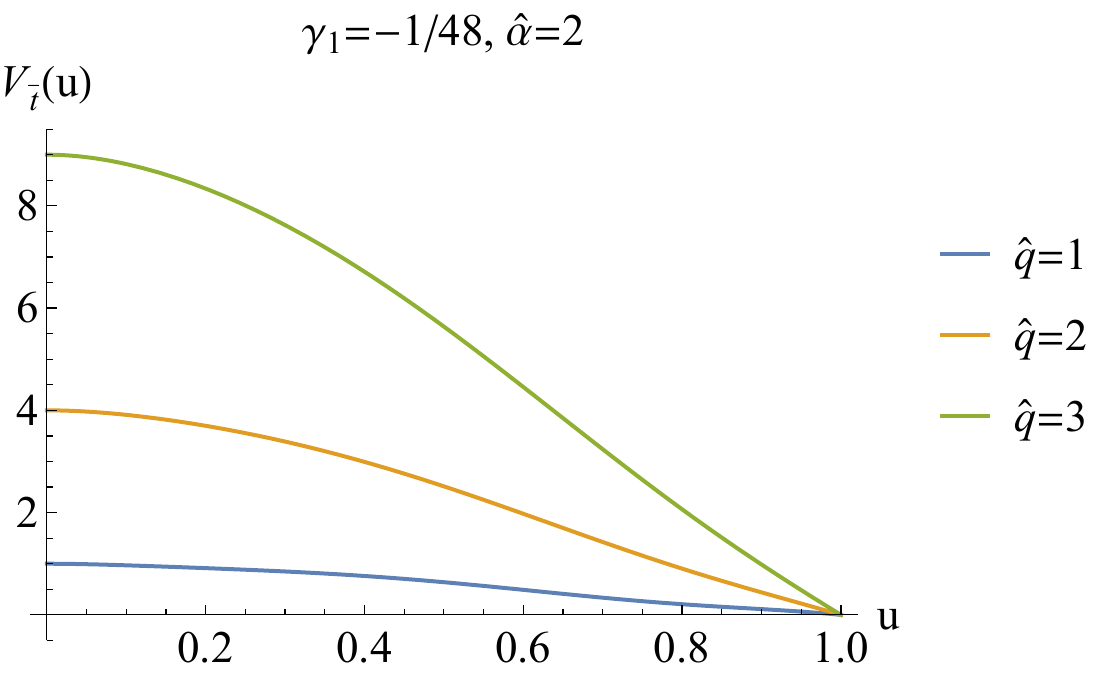}\ \\
\caption{\label{fig-Vt} The potentials $V_{\bar{t}}(u)$ with different $\gamma_1$ and $\hat{\alpha}$ at some finite momentum.}}
\end{figure}

After examining the instabilities for small and large momentum limit, we turn to the case for some finite momentum.
For this aim, we show the potentials $V_{\bar{t}}(u)$ for different $\gamma_1$ and $\hat{\alpha}$ at some finite momentum in FIG.\ref{fig-Vt}.
It is obvious that the potential is always positive, which indicates that no unstable modes appear even for the finite momentum.

Therefore, we conclude that the region $\gamma_1\in(-\infty,+\infty)$ is still physically viable for EA-AdS geometry.
Combining with the constraint from conductivity, we have $\gamma_1\in(-\infty,1/48]$.

\section{Conductivity}\label{sec-optical-conductivity}

In this section, we study the optical conductivity by
solving the second equation in (\ref{Ma-Ay}) for $\hat{q}=0$ with ingoing boundary condition at horizon.
And then it can be read off in terms of
\fa
\sigma(\omega)=\frac{\partial_uA_y(u,\hat{\omega},\hat{q}=0)}{i\omega A_y(u,\hat{\omega},\hat{q}=0)}\,.
\label{con-def}
\ffa

\begin{figure}
\center{
\includegraphics[scale=0.6]{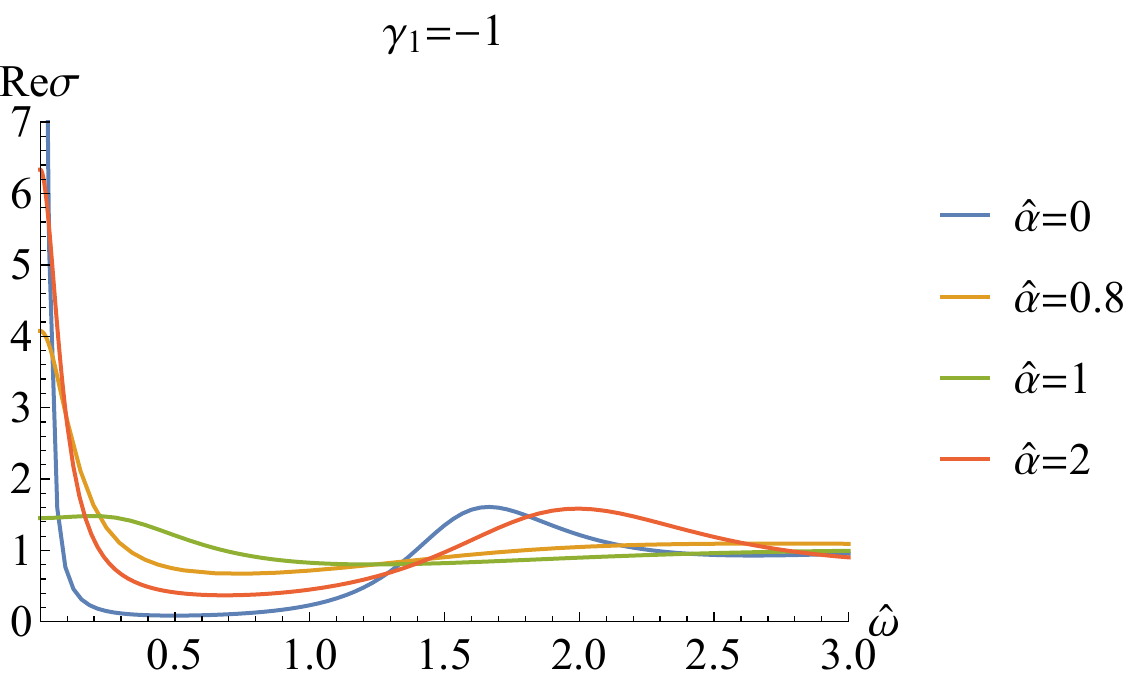}\ \hspace{0.8cm}
\includegraphics[scale=0.6]{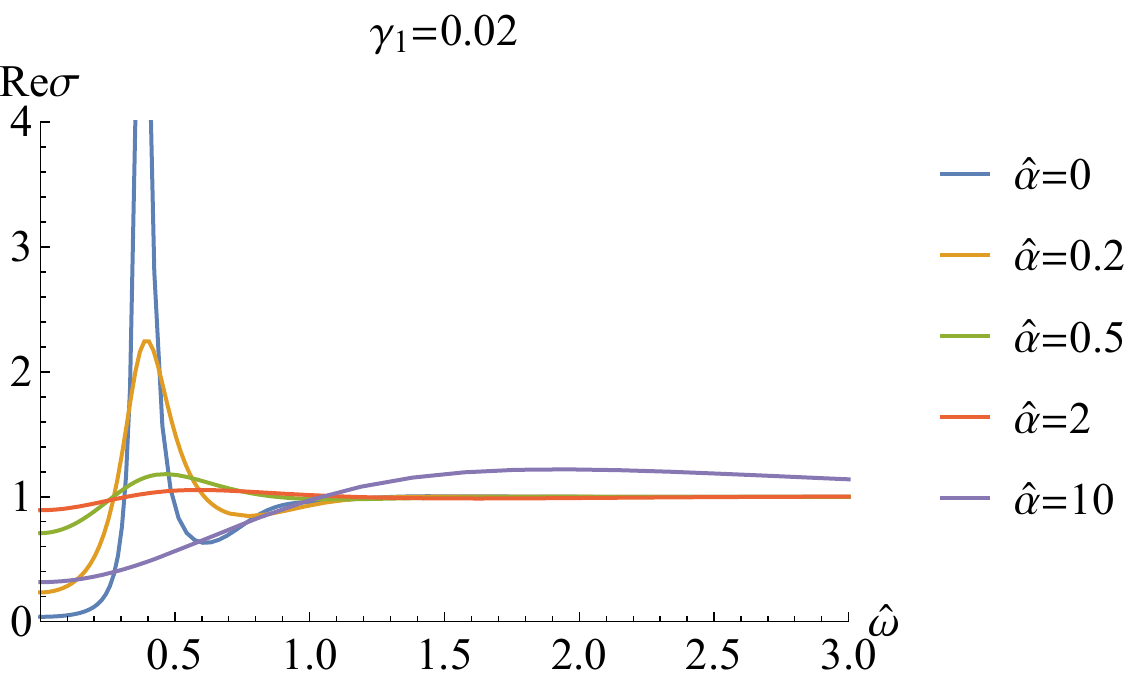}\ \\
\caption{\label{fig-con-v1} Left plot: The optical conductivity with $\gamma_1=-1$ for different $\hat{\alpha}$.
Right plot: The optical conductivity with $\gamma_1=0.02$ for different $\hat{\alpha}$.}}
\end{figure}
\begin{figure}
\center{
\includegraphics[scale=0.6]{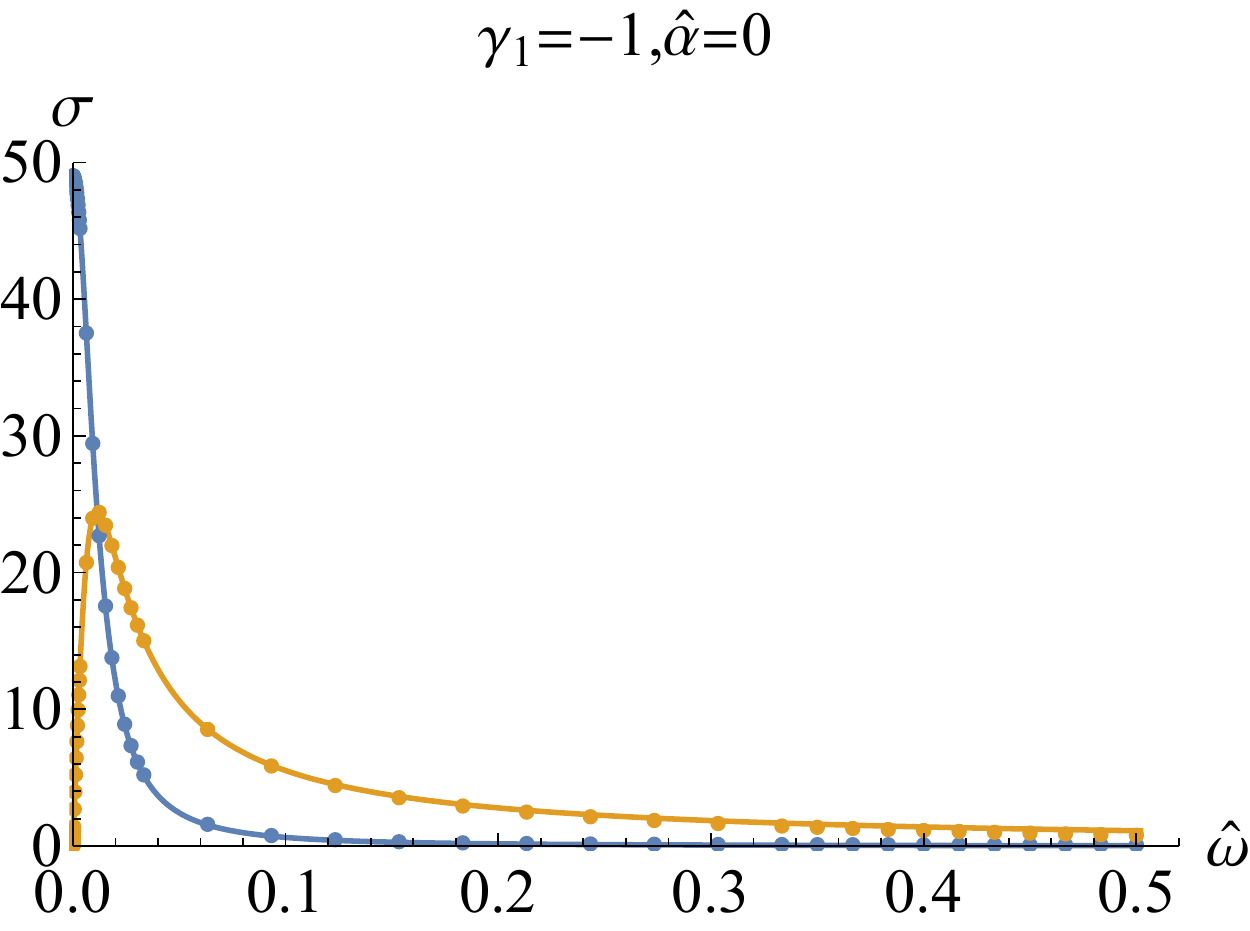}\ \hspace{0.8cm}
\includegraphics[scale=0.6]{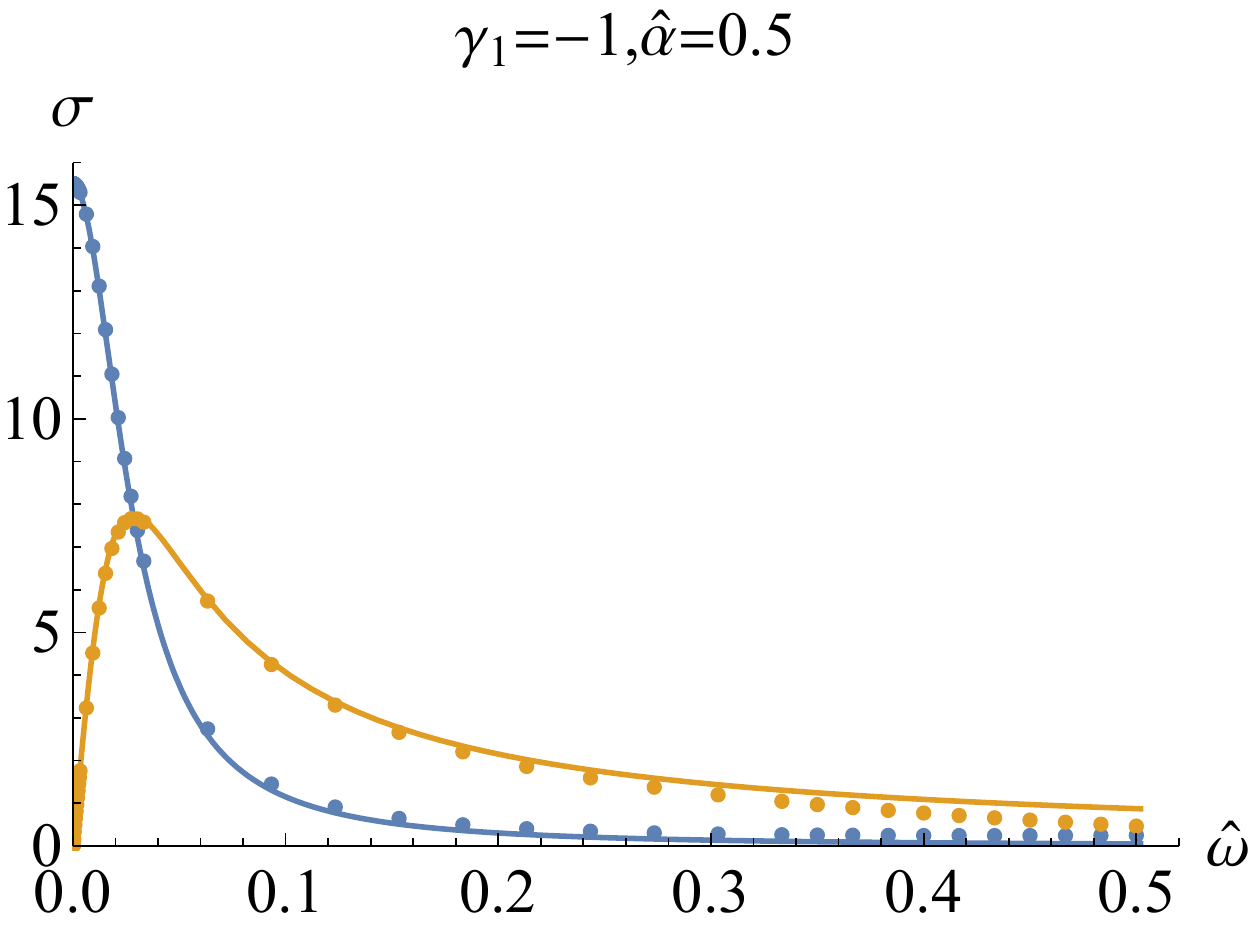}\ \\
\caption{\label{fig-fit-v1} The low frequency behavior of the optical conductivity.
The real line is fitted with the standard Drude formula (\ref{Drude}).}}
\end{figure}
\begin{figure}
\center{
\includegraphics[scale=0.6]{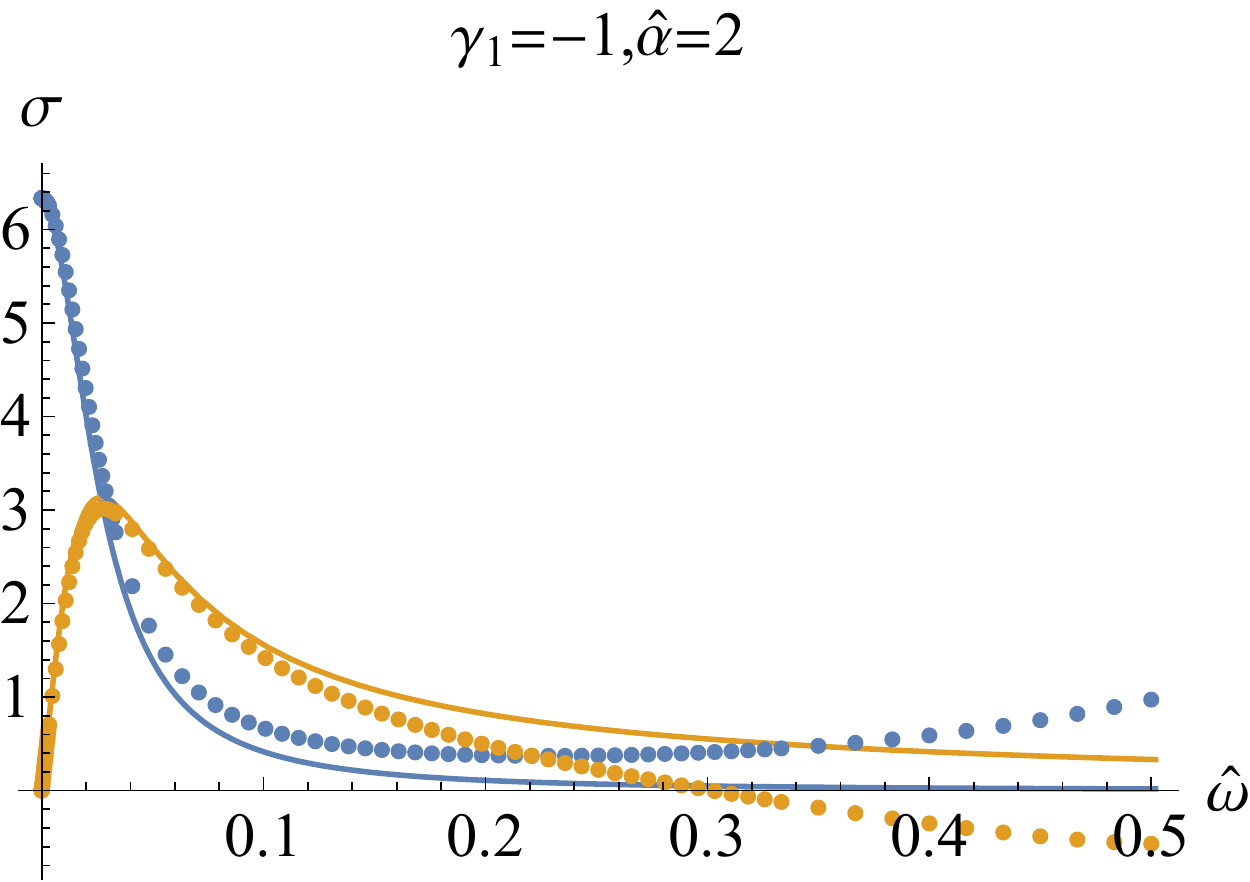}\ \\
\caption{\label{fig-fit-v2} The low frequency behavior of the optical conductivity.
The real line is fitted with the standard Drude formula (\ref{Drude}).}}
\end{figure}

As has been shown in \cite{Witczak-Krempa:2013aea}, for $\gamma_1=-1$ and $\hat{\alpha}=0$,
the optical conductivity $\sigma(\omega)$ exhibits a sharp Drude-like peak at small frequency,
then a gap and eventually rises and saturates to the value $\sigma=1$ at large frequency (left plot in FIG.\ref{fig-con-v1}).
While for $\gamma_1=0.02$ and $\hat{\alpha}=0$, it displays a hard-gap-like behavior at small frequency,
then a pronounced peak before saturating (right plot in FIG.\ref{fig-con-v1}).
The holographic result is similar with that
of the $O(N)$ $NL\sigma M$ in the large-$N$ limit \cite{Damle:1997rxu} as pointed out in \cite{Witczak-Krempa:2013aea}.

Now we study the effect of the homogeneous disorder.
For small $\hat{\alpha}\ll 1$, we find that the transport is well described by the standard Drude formula (see FIG.\ref{fig-fit-v1})
\fa
\sigma(\hat{\omega})=\frac{K\tau}{1-i\hat{\omega}\tau}\,.
\label{Drude}
\ffa
TABLE \ref{table-Drude} shows some examples of the relaxation time $\tau$
for different $\hat{\alpha}$. It decreased with the increase of $\hat{\alpha}$,
indicating that the dissipation of this system gradually becomes obvious.
With $\hat{\alpha}$ further increasing before reaching $\hat{\alpha}=2/\sqrt{3}$,
the Drude-like peak disappears, indicating a crossover from a coherent to an incoherent metallic phase.
After $\hat{\alpha}$ is beyond $\hat{\alpha}=2/\sqrt{3}$,
with it increasing further, a peak exhibits again (left plot in FIG.\ref{fig-con-v1}).
It is different from that of $4$ derivatives showing in \cite{Wu:2016jjd}, in which
the low frequency conductivity of the holographic system with $\gamma>0$ but other coupling parameter vanishing
exhibits a dip as $\hat{\alpha}$ becomes larger.
But the peak is not well fitted by the standard Drude formula (see FIG.\ref{fig-fit-v2}).
It indicates that we need new formula to fit it and we shall further explore it in future.
While for $\gamma_1=0.02$ and $\hat{\alpha}\neq 0$, a soft gap exhibits at low frequency replacing the hard gap
because of the effect of homogeneous disorder.
And the sharp resonant peak at medium frequency for $\hat{\alpha}=0$
gradually decreases with the increase of $\hat{\alpha}$.
We expect that the $O(N)$ $NL\sigma M$ in the large-$N$ limit \cite{Damle:1997rxu}
exhibits some similar phenomena when some effects of disorder are introduced.
We shall explore it in future.

\begin{widetext}
\begin{table}[ht]
\begin{center}
\begin{tabular}{|c|c|c|c|c|c|c|}
         \hline
~$\hat{\alpha}$~ &~$0$~&~$0.01$~&~$0.1$~&~$0.5$~
          \\
        \hline
~$\tau$~ & ~$87.62$~&~$86.64$~&~$83.67$~&~$35.27$~
          \\
        \hline
\end{tabular}
\caption{\label{table-Drude} The relaxation time $\tau$ of the holographic system with $\gamma_1=-1$
fitted by the standard Drude formula (\ref{Drude}) for different $\hat{\alpha}$.}
\end{center}
\end{table}
\end{widetext}

\section{Conclusion and discussion}\label{sec-conclusion}

In this letter, we mainly study the charge response from higher derivatives, in particular the $6$ derivatives,
over the homogeneous disorder background.
In \cite{Witczak-Krempa:2013aea}, it has been shown that the holographic result from higher derivatives is similar with that
of the $O(N)$ $NL\sigma M$ in the large-$N$ limit \cite{Damle:1997rxu}.
When the homogeneous disorder effect by axions is introduced into this holographic system with $\gamma_1<0$ (small enough, for example $\gamma_1=-1$),
we find that before $\hat{\alpha}$ reaching $\hat{\alpha}=2/\sqrt{3}$,
there is a crossover from a coherent to incoherent metallic phase with $\hat{\alpha}$ increasing.
As $\hat{\alpha}$ is beyond $\hat{\alpha}=2/\sqrt{3}$ and further amplified,
a peak exhibits again at low frequency.
But it cannot be well fitted by the standard Drude formula
and call for new formula.
While the holographic system with the limit of $\gamma_1\rightarrow 1/48$
is an insulator with a hard-gap-like at low frequency
and a pronounced peak emerging at medium frequency.
But the homogeneous disorder effect drives the hard-gap-like at low frequency into
the soft gap and suppresses the pronounced peak at medium frequency.
We expect that when some effects of disorder are introduced into the $O(N)$ $NL\sigma M$ in the large-$N$ limit,
some similar phenomena can be observed, which deserves further studying.

It is also interesting and illuminating to compare the main properties of the conductivity
from $6$ derivatives ($\gamma=0$ and $\gamma_1\neq0$) here
with that from $4$ derivatives ($\gamma\neq 0$ and $\gamma_1=0$) in our previous work \cite{Wu:2016jjd}.
First, for the case of $4$ derivatives,
since the causality results to a bound on $\gamma$,
the transport is incoherent for any $\hat{\alpha}$.
While for the case of $6$ derivatives,
there is coherent charge transport with long-lived excitations for $\gamma_1<0$ and small $\hat{\alpha}$.
Second, there is a dip in optical conductivity with $\gamma<0$ and $\gamma_1=0$ for small $\hat{\alpha}$,
which is similar with the excitation of vortices.
But for the case with $\gamma_1>0$ and $\gamma=0$,
the optical conductivity exhibits a hard-gap-like or soft gap at low frequency
and a resonant peak at medium frequency.
Third, for the case of $4$ derivatives,
the homogeneous disorder effect drives the peak (dip) in the low frequency optical conductivity into its contrary.
But for the case of $6$ derivatives,
the homogeneous disorder effect cannot drive the peak (gap) into its contrary.
It is an interesting phenomenon and deserves further exploring for other higher derivative terms.

There are lots of open questions deserving further exploration.
For example, we can study the charge responses
from higher derivatives with homogeneous disorder at full momentum and energy spaces.
Also we can test the robustness of the phenomena observed here
when the disorder is introduced by the other way, for instance \cite{Grozdanov:2015qia,Donos:2013eha,Donos:2014uba,Ling:2015exa,Ling:2015epa,Vegh:2013sk,Blake:2013bqa,Donos:2012js,Kim:2014bza,Davison:2014lua,Gouteraux:2016wxj,Ling:2016lis}.

\begin{acknowledgments}

This work is supported by the Natural Science Foundation of China under
Grant No.11305018, by Natural Science Foundation of Liaoning Province under
Grant No.201602013,
and by the grant (No.14DZ2260700) from the Opening Project of Shanghai Key Laboratory
of High Temperature Superconductors.

\end{acknowledgments}

\begin{appendix}

\end{appendix}


\begin{thebibliography}{99}


\bibitem{Sachdev:QPT}
S. Sachdev, ``Quantum Phase Transitions'', Cambridge University Press, England, 2nd edition, (2011).


\bibitem{Damle:1997rxu}
  K.~Damle and S.~Sachdev,
  ``Nonzero-temperature transport near quantum critical points,''
  Phys.\ Rev.\ B {\bf 56}, no. 14, 8714 (1997)
  [cond-mat/9705206 [cond-mat.str-el]].

\bibitem{WitczakKrempa:2012um}
  W.~Witczak-Krempa, P.~Ghaemi, T.~Senthil and Y.~B.~Kim,
  ``Universal transport near a quantum critical Mott transition in two dimensions,''
  Phys.\ Rev.\ B {\bf 86}, 245102 (2012)
  [arXiv:1206.3309 [cond-mat.str-el]].


\bibitem{Cha:1991}
  Min-Chul Cha, Matthew P. A. Fisher, S. M. Girvin, Mats Wallin, and A. Peter Young,
  ``Universal conductivity of two-dimensional films at the superconductor-insulator transition,''
  Phys. Rev. B {\bf 44}, 6883-6902 (1991).

\bibitem{Smakov:2005}
 J. Smakov, E. Sorensen,
 ``Universal Scaling of the Conductivity at the Superfluid-Insulator Phase Transition,''
 Physical Review Letters {\bf 95}, 180603 (2005), [arXiv:cond-mat/0509671].

\bibitem{Chen:2013ppa}
  K.~Chen, L.~Liu, Y.~Deng, L.~Pollet and N.~Prokof'ev,
  ``Universal Conductivity in a Two-Dimensional Superfluid-to-Insulator Quantum Critical System,''
  Phys.\ Rev.\ Lett.\  {\bf 112}, no. 3, 030402 (2014)
  [arXiv:1309.5635 [cond-mat.str-el]].

\bibitem{Gazit:2013hga}
  S.~Gazit, D.~Podolsky, A.~Auerbach and D.~P.~Arovas,
  ``Dynamics and Conductivity Near Quantum Criticality,''
  Phys.\ Rev.\ B {\bf 88}, 235108 (2013)
  [arXiv:1309.1765 [cond-mat.str-el]].

\bibitem{Gazit:2014}
S. Gazit, D. Podolsky, and A. Auerbach,
``Critical Capacitance and Charge-Vortex Duality Near the Superfluid-to-Insulator Transition,'' Phys. Rev. Lett.
{\bf 113}, 240601 (2014), [arXiv:1407.1055 [cond-mat.str-el]].

\bibitem{Witczak-Krempa:2015jca}
  W.~Witczak-Krempa and J.~Maciejko,
  ``Optical conductivity of topological surface states with emergent supersymmetry,''
  Phys.\ Rev.\ Lett.\  {\bf 116}, no. 10, 100402 (2016)
  [arXiv:1510.06397 [cond-mat.str-el]].

\bibitem{Lucas:2016fju}
  A.~Lucas, S.~Gazit, D.~Podolsky and W.~Witczak-Krempa,
  ``Dynamical response near quantum critical points,''
  arXiv:1608.02586 [cond-mat.str-el].


\bibitem{Maldacena:1997re}
  J.~M.~Maldacena,
  ``The Large N limit of superconformal field theories and supergravity,''
  Int.\ J.\ Theor.\ Phys.\  {\bf 38}, 1113 (1999)
  [Adv.\ Theor.\ Math.\ Phys.\  {\bf 2}, 231 (1998)]
  [hep-th/9711200].

\bibitem{Gubser:1998bc}
  S.~S.~Gubser, I.~R.~Klebanov and A.~M.~Polyakov,
  ``Gauge theory correlators from noncritical string theory,''
  Phys.\ Lett.\ B {\bf 428}, 105 (1998)
  [hep-th/9802109].

\bibitem{Witten:1998qj}
  E.~Witten,
  ``Anti-de Sitter space and holography,''
  Adv.\ Theor.\ Math.\ Phys.\  {\bf 2}, 253 (1998)
  [hep-th/9802150].

\bibitem{Aharony:1999ti}
  O.~Aharony, S.~S.~Gubser, J.~M.~Maldacena, H.~Ooguri and Y.~Oz,
  ``Large N field theories, string theory and gravity,''
  Phys.\ Rept.\  {\bf 323}, 183 (2000)
  [hep-th/9905111].

\bibitem{Myers:2010pk}
  R.~C.~Myers, S.~Sachdev and A.~Singh,
  ``Holographic Quantum Critical Transport without Self-Duality,''
  Phys.\ Rev.\ D {\bf 83}, 066017 (2011)
  [arXiv:1010.0443 [hep-th]].

\bibitem{Sachdev:2011wg}
  S.~Sachdev,
  ``What can gauge-gravity duality teach us about condensed matter physics?,''
  Ann.\ Rev.\ Condensed Matter Phys.\  {\bf 3}, 9 (2012)
  [arXiv:1108.1197 [cond-mat.str-el]].

\bibitem{Hartnoll:2016apf}
  S.~A.~Hartnoll, A.~Lucas and S.~Sachdev,
  ``Holographic quantum matter,''
  arXiv:1612.07324 [hep-th].

\bibitem{Witczak-Krempa:2013aea}
  W.~Witczak-Krempa,
  ``Quantum critical charge response from higher derivatives in holography,''
  Phys.\ Rev.\ B {\bf 89}, no. 16, 161114 (2014)
  [arXiv:1312.3334 [cond-mat.str-el]].


\bibitem{Ritz:2008kh}
  A.~Ritz and J.~Ward,
  ``Weyl corrections to holographic conductivity,''
  Phys.\ Rev.\ D {\bf 79}, 066003 (2009)
  [arXiv:0811.4195 [hep-th]].

\bibitem{Wu:2016jjd}
  J.~P.~Wu,
  ``Momentum dissipation and holographic transport without self-duality,''
  arXiv:1609.04729 [hep-th].

\bibitem{Andrade:2013gsa}
  T.~Andrade and B.~Withers,
  ``A simple holographic model of momentum relaxation,''
  JHEP {\bf 1405}, 101 (2014)
  [arXiv:1311.5157 [hep-th]].

\bibitem{Grozdanov:2015qia}
  S.~Grozdanov, A.~Lucas, S.~Sachdev and K.~Schalm,
  ``Absence of disorder-driven metal-insulator transitions in simple holographic models,''
  Phys.\ Rev.\ Lett.\  {\bf 115}, no. 22, 221601 (2015)
  [arXiv:1507.00003 [hep-th]].


\bibitem{Donos:2012js}
  A.~Donos and S.~A.~Hartnoll,
  ``Interaction-driven localization in holography,''
  Nature Phys.\  {\bf 9}, 649 (2013)
  [arXiv:1212.2998].

\bibitem{Donos:2013eha}
  A.~Donos and J.~P.~Gauntlett,
  ``Holographic Q-lattices,''
  JHEP {\bf 1404}, 040 (2014)
  [arXiv:1311.3292 [hep-th]].

\bibitem{Donos:2014uba}
  A.~Donos and J.~P.~Gauntlett,
  ``Novel metals and insulators from holography,''
  JHEP {\bf 1406}, 007 (2014)
  [arXiv:1401.5077 [hep-th]].

\bibitem{Vegh:2013sk}
  D.~Vegh,
  ``Holography without translational symmetry,''
  arXiv:1301.0537 [hep-th].

\bibitem{Blake:2013bqa}
  M.~Blake and D.~Tong,
  ``Universal Resistivity from Holographic Massive Gravity,''
  Phys.\ Rev.\ D {\bf 88}, no. 10, 106004 (2013)
  [arXiv:1308.4970 [hep-th]].

\bibitem{Gouteraux:2016wxj}
  B.~Gouteraux, E.~Kiritsis and W.~J.~Li,
  ``Effective holographic theories of momentum relaxation and violation of conductivity bound,''
  JHEP {\bf 1604}, 122 (2016)
  [arXiv:1602.01067 [hep-th]].

\bibitem{Ling:2016lis}
  Y.~Ling and X.~Zheng,
  ``Holographic superconductor with momentum relaxation and Weyl correction,''
  arXiv:1609.09717 [hep-th].


\bibitem{Kim:2014bza}
  K.~Y.~Kim, K.~K.~Kim, Y.~Seo and S.~J.~Sin,
  ``Coherent/incoherent metal transition in a holographic model,''
  JHEP {\bf 1412}, 170 (2014)
  [arXiv:1409.8346 [hep-th]].

\bibitem{Davison:2014lua}
  R.~A.~Davison and B.~Gout¨¦raux,
  ``Momentum dissipation and effective theories of coherent and incoherent transport,''
  JHEP {\bf 1501}, 039 (2015)
  [arXiv:1411.1062 [hep-th]].

\bibitem{Ling:2015epa}
  Y.~Ling, P.~Liu, C.~Niu and J.~P.~Wu,
  ``Building a doped Mott system by holography,''
  Phys.\ Rev.\ D {\bf 92}, no. 8, 086003 (2015)
  [arXiv:1507.02514 [hep-th]].

\bibitem{Ling:2015exa}
  Y.~Ling, P.~Liu and J.~P.~Wu,
  ``A novel insulator by holographic Q-lattices,''
  JHEP {\bf 1602}, 075 (2016)
  [arXiv:1510.05456 [hep-th]].



\bibitem{WitczakKrempa:2012gn}
  W.~Witczak-Krempa and S.~Sachdev,
  ``The quasi-normal modes of quantum criticality,''
  Phys.\ Rev.\ B {\bf 86}, 235115 (2012)
  [arXiv:1210.4166 [cond-mat.str-el]].

\bibitem{WitczakKrempa:2013ht}
  W.~Witczak-Krempa and S.~Sachdev,
  ``Dispersing quasinormal modes in 2+1 dimensional conformal field theories,''
  Phys.\ Rev.\ B {\bf 87}, 155149 (2013)
  [arXiv:1302.0847 [cond-mat.str-el]].

\bibitem{Witczak-Krempa:2013nua}
  W.~Witczak-Krempa, E.~S.~S{\o}rensen and S.~Sachdev,
  ``The dynamics of quantum criticality via Quantum Monte Carlo and holography,''
  Nature Phys.\  {\bf 10}, 361 (2014)
  [arXiv:1309.2941 [cond-mat.str-el]].

\bibitem{Katz:2014rla}
  E.~Katz, S.~Sachdev, E.~S.~S{\o}rensen and W.~Witczak-Krempa,
  ``Conformal field theories at nonzero temperature: Operator product expansions, Monte Carlo, and holography,''
  Phys.\ Rev.\ B {\bf 90}, no. 24, 245109 (2014)
  [arXiv:1409.3841 [cond-mat.str-el]].

\bibitem{Bai:2013tfa}
  S.~Bai and D.~W.~Pang,
  ``Holographic charge transport in 2+1 dimensions at finite $N$,''
  Int.\ J.\ Mod.\ Phys.\ A {\bf 29}, 1450061 (2014)
  [arXiv:1312.3351 [hep-th]].

\bibitem{Wu:2010vr}
  J.~P.~Wu, Y.~Cao, X.~M.~Kuang and W.~J.~Li,
  ``The 3+1 holographic superconductor with Weyl corrections,''
  Phys.\ Lett.\ B {\bf 697}, 153 (2011)
  [arXiv:1010.1929 [hep-th]].

\bibitem{Ma:2011zze}
  D.~Z.~Ma, Y.~Cao and J.~P.~Wu,
  ``The Stuckelberg holographic superconductors with Weyl corrections,''
  Phys.\ Lett.\ B {\bf 704}, 604 (2011)
  [arXiv:1201.2486 [hep-th]].
\bibitem{Momeni:2011ca}
  D.~Momeni and M.~R.~Setare,
  ``A note on holographic superconductors with Weyl Corrections,''
  Mod.\ Phys.\ Lett.\ A {\bf 26}, 2889 (2011)
  [arXiv:1106.0431 [physics.gen-ph]].
\bibitem{Momeni:2012ab}
  D.~Momeni, N.~Majd and R.~Myrzakulov,
  ``p-wave holographic superconductors with Weyl corrections,''
  Europhys.\ Lett.\  {\bf 97}, 61001 (2012)
  [arXiv:1204.1246 [hep-th]].
\bibitem{Zhao:2012kp}
  Z.~Zhao, Q.~Pan and J.~Jing,
  ``Holographic insulator/superconductor phase transition with Weyl corrections,''
  Phys.\ Lett.\ B {\bf 719}, 440 (2013)
  [arXiv:1212.3062].
\bibitem{Momeni:2013fma}
  D.~Momeni, R.~Myrzakulov and M.~Raza,
  ``Holographic superconductors with Weyl Corrections via gauge/gravity duality,''
  Int.\ J.\ Mod.\ Phys.\ A {\bf 28}, 1350096 (2013)
  [arXiv:1307.8348 [hep-th]].
\bibitem{Momeni:2014efa}
  D.~Momeni, M.~Raza and R.~Myrzakulov,
  ``Holographic superconductors with Weyl corrections,''
  Int.\ J.\ Geom.\ Meth.\ Mod.\ Phys.\  {\bf 13}, 1550131 (2016)
  [arXiv:1410.8379 [hep-th]].
\bibitem{Zhang:2015eea}
  L.~Zhang, Q.~Pan and J.~Jing,
  ``Holographic p-wave superconductor models with Weyl corrections,''
  Phys.\ Lett.\ B {\bf 743}, 104 (2015)
  [arXiv:1502.05635 [hep-th]].

\bibitem{Mansoori:2016zbp}
  S.~A.~H.~Mansoori, B.~Mirza, A.~Mokhtari, F.~L.~Dezaki and Z.~Sherkatghanad,
  ``Weyl holographic superconductor in the Lifshitz black hole background,''
  JHEP {\bf 1607}, 111 (2016)
  [arXiv:1602.07245 [hep-th]].

\bibitem{Ling:2016dck}
  Y.~Ling, P.~Liu, J.~P.~Wu and Z.~Zhou,
  ``Holographic Metal-Insulator Transition in Higher Derivative Gravity,''
  Phys.\ Lett.\ B {\bf 766}, 41 (2017)
  [arXiv:1606.07866 [hep-th]].

\bibitem{Myers:2007we}
  R.~C.~Myers, A.~O.~Starinets and R.~M.~Thomson,
  ``Holographic spectral functions and diffusion constants for fundamental matter,''
  JHEP {\bf 0711}, 091 (2007)
  [arXiv:0706.0162 [hep-th]].



\end{thebibliography}
\end{document}